\documentclass[runningheads]{llncs}
\usepackage[T1]{fontenc}
\usepackage{graphicx}
 \usepackage{amsmath}
\begin{document}
\title{Discovering Frequent Closed Embedded Sub-DAGs in Spatio-Temporal Event Data}
\titlerunning{Frequent Closed Sub-DAGs in Spatio-Temporal Events}
%
\author{Piotr S. Maciąg\orcidID{0000-0001-5486-7927}}
\authorrunning{Piotr Maciąg}
%
\institute{Institute of Computer Science, Warsaw University of Technology, Nowowiejska 15/19, 00-661, Warsaw
\email{piotr.maciag@pw.edu.pl}}
\maketitle              
\begin{abstract}
We propose a novel approach to mine patterns in spatio-temporal event data based on discovering frequent closed embedded sub-Directed Acyclic Graphs (DAGs). In our method, event instances are represented as nodes labelled by event types, while edges capture spatio-temporal following relationships. We formally define the considered class of patterns and provide the rationale for focusing on closed sub-DAGs as compact and non-redundant representations of recurring interaction patterns. We implement the DigDag algorithm for mining such patterns and experimentally compare its efficiency with two related approaches: propagation pattern mining using the SLEUTH algorithm and Cascading Spatio-Temporal Pattern mining using the CSTPM algorithm. The experimental results demonstrate that our approach is substantially more efficient while operating under comparable parameter settings. Finally, we present a qualitative analysis of selected discovered patterns.
\keywords{spatio-temporal pattern mining  \and  frequent closed sub-DAGs \and crime event analysis.}
\end{abstract}
\section{Introduction}

Discovering various spatio-temporal patterns has long been a topic attracting the attention of both researchers and practitioners. The literature in this area is extensive, as the nature of spatio-temporal data is inherently complex and multimodal. For example, researchers analysing trajectories of animal flocks are interested in different types of patterns than those attempting to extract knowledge from datasets of epidemic incidents, where only approximate spatial and temporal occurrences are available for each event \cite{ref1284:Atluri2018,ref1284:Li2014}.

In this work, we focus on discovering spatio-temporal patterns from instances of crime events occurring in municipal areas. We assume that each event instance is described by a unique identifier, a geographical location, an occurrence time, and an event type. Our objective is to uncover mutual relationships between different types of crimes in the dataset. For example, we aim to investigate whether the occurrence of one crime type at a given location is followed by subsequent occurrences of other crime types within the same spatial area. 

To this end, we apply closed frequent embedded sub-Directed Acyclic Graph (DAG) mining. In our approach, the nodes of a DAG correspond to individual crime event instances, and their labels represent event types. The edges between nodes express the spatio-temporal following relationship between event instances. For example, they may capture the fact that the occurrence of an event instance of one crime type is followed by subsequent occurrences of other crime types within a given spatial and temporal neighbourhood.

In this context, a frequent sub-DAG can be intuitively understood as a subgraph pattern that occurs sufficiently often across the constructed DAGs and therefore represents a recurring spatio-temporal interaction among different crime types. By focusing on closed patterns, we eliminate redundant substructures and retain only the most informative and non-redundant representations of such relationships.


\textbf{Contributions.} 

The contributions of this article are as follows:
\begin{itemize}
    \item We propose an approach for discovering frequent closed (embedded) sub-DAGs in spatio-temporal event data. We introduce the necessary definitions in Section~\ref{sec:BasicNotions} and provide the rationale for focusing on this class of patterns. Furthermore, we implement the DigDag algorithm to discover frequent closed sub-DAGs from a set of crime event instances.

    \item We experimentally compare the efficiency of discovering the proposed type of patterns with the efficiency of discovering related propagation patterns introduced in the literature~\cite{Moosavi2019-ShortLongSTPatterns}. As demonstrated in our experiments, mining the proposed patterns using the DigDag algorithm is significantly more efficient and technically more robust to varying parameters' setups than discovering propagation patterns with the SLEUTH algorithm~\cite{Moosavi2019-ShortLongSTPatterns}.
    

    \item We experimentally compare the efficiency of discovering the proposed patterns with the efficiency of mining CSTPs~\cite{ref1284:Mohan2012} using the CSTPM algorithm\footnote{We implemented all three algorithms evaluated in this study: DigDag, SLEUTH, and CSTPM in Python.}. Our experimental results indicate that the proposed approach is substantially more efficient than the CSTPM algorithm.

    \item We conduct qualitative analysis of some of the discovered patterns.
\end{itemize}


\section{Related Work}
\label{sec:RelatedWork}

In general, the problem of discovering spatio-temporal patterns from event data across various domains, such as crime, weather, transportation, and air pollution, has been extensively investigated in the related literature:
\begin{itemize}
    \item Spatio-temporal Sequential Patterns (STSPs), introduced in~\cite{ref1284:Huang2008} and further extended in subsequent works~\cite{ref1284:Maciag2019,Maciag2025-Geoinformatica}, are defined as sequences of event types. A sequence $\overrightarrow{s}$ of event types $F_1 \rightarrow F_2 \rightarrow \dots \rightarrow F_k$ expresses an occurrence dependency among these types, such that, within their spatio-temporal neighbourhoods, instances of type $F_1$ either attract or repel occurrences of type $F_2$. In turn, instances of type $F_2$ attract or repel occurrences of the subsequent event type, and so on, up to type $F_k$.

    The strength of the relationship between consecutive event types, as proposed in~\cite{ref1284:Huang2008}, is quantified using the Participation Ratio (PR) measure, while the overall significance of a sequence $\overrightarrow{s}$ is evaluated using the Sequence Index (SI), defined as the minimum of all corresponding PR values. STSPs have been successfully applied in domains such as transportation, air pollution monitoring, and weather analysis.

    \item Mohan et al.~\cite{ref1284:Mohan2012} introduced Cascading Spatio-Temporal Patterns (CSTPs), which are defined as cascades (i.e., partially ordered sets) of event types. In contrast to STSPs, CSTPs provide more expressive information about the relationships among occurrences of different event types. For example, they can capture situations in which occurrences of one event type subsequently trigger the mutual occurrence of two or more different event types within the same spatial area.

    To assess the significance of CSTPs, \cite{ref1284:Mohan2012} proposed the Cascading Participation Index (CPI) as the corresponding interestingness measure. 

    \item Moosavi et al.~\cite{Moosavi2019-ShortLongSTPatterns} proposed propagation patterns. Propagation patterns are defined as frequent embedded subtrees occurring within a forest, where each tree consists of labelled nodes representing event instances (with node labels corresponding to event types). The edges between nodes are constructed based on a spatio-temporal following relationship. 
\end{itemize}

In contrast to the approach of Moosavi et al.~\cite{Moosavi2019-ShortLongSTPatterns}, which focuses on mining \textit{frequent subtrees} (aka. propagation patterns), we concentrate on mining \textit{frequent (closed) sub-DAGs}. We argue that the DAG representation is more natural for modelling relationships among different event types in a spatio-temporal event dataset.

\begin{figure}
    \includegraphics[width=\textwidth]{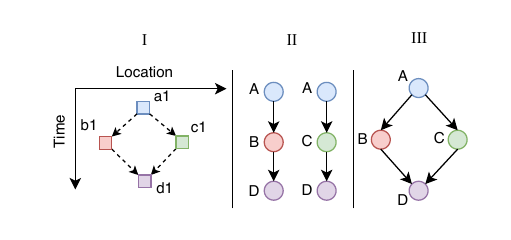}
    \caption{Four event instances of four event types $(A, B, C, D)$ connected using spatio-temporal following relationship; corresponding trees and a DAG of event types.} 
\label{fig:Comparison}
\end{figure}

For example, consider a simple dataset consisting of four event instances $a_1, b_1, c_1, d_1$, as illustrated in Fig.~\ref{fig:Comparison}.I. The edges between event instances denote the spatio-temporal following relationship. In Fig.~\ref{fig:Comparison}.II, the dataset is transformed into an example forest of trees, whereas in Fig.~\ref{fig:Comparison}.III it is represented as a DAG. 

Note that the tree representation shown in Fig.~\ref{fig:Comparison}.II fails to capture the fact that the occurrence of an event instance of type $D$ may be induced only by the simultaneous prior occurrences of events of types $B$ and $C$.

Similarly to our approach, the CSTPs proposed by Mohan et al.~\cite{ref1284:Mohan2012} also require DAGs as input data. However, as demonstrated in our experiments, discovering CSTP patterns using the method of~\cite{ref1284:Mohan2012} can be substantially more computationally expensive, even when comparable parameter settings are used in both approaches.

Some other studies have focused on discovering concise representations of spatio-temporal patterns, including closed patterns \cite{ref1284:Maciag2019-Kes}, constricted patterns \cite{Maciag2023-Constr}, and targeted patterns \cite{Maciag2025-Geoinformatica}. The goal of these approaches is to return a smaller, non-redundant set of patterns, typically by eliminating redundancy, while largely preserving the informativeness of the original pattern set. Other research—though less central to this research—has investigated mining spatio-temporal sequential patterns from more complex objects, such as regions \cite{Aydin2016-event-sequences,Aydin2020-Geoinformatica} and trajectories \cite{Kong2018-BigTrajectoryData}. For a broader overview of research directions in spatio-temporal data mining, we refer the reader to \cite{Hamdi2022}.

\section{The Proposed Approach}
\label{sec:BasicNotions}

Formally, we assume that our input dataset $\mathbf{D}$ consists of events instances $i \in \mathbf{D}$, each of which has a location (e.g., longitude and latitude) $i.location$, an occurrence time $i.time$ and a type $i.F$. The set of all event (crime) types is denoted as $\mathbf{F}$. 

\begin{definition}[Spatio-temporal following relationship]
    For two event instances $i_1$ and $i_2$, we say that $i_2$ follows $i_1$ ($i_1 \prec i_2$), if: 1. the distance between $i_1.location$ and $i_2.location$, is $\leq R$ and 2. $(i_2.time - i_1.time) \in (0, T]$, where $R$ and $T$ are the user-given parameters. $R$ is a spatial radius threshold and $T$ is a time threshold. 
\end{definition}

\begin{figure}
    \includegraphics[width=\textwidth]{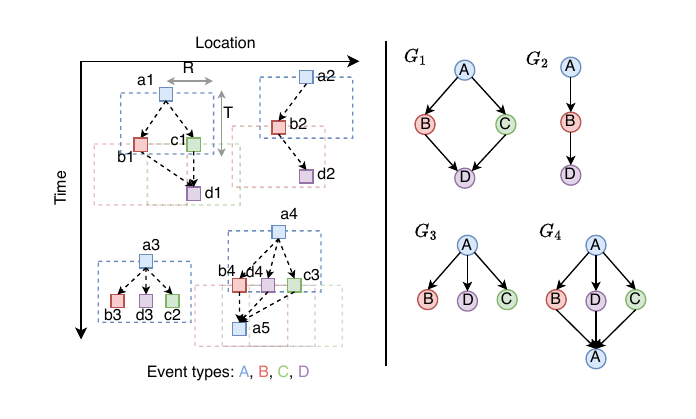}
    \caption{An example dataset $\mathbf{D}$ of four event types $\mathbf{F}$. The edges between event instances are created according to Definition~1 for shown parameters $R$, $T$.} 
\label{fig:Dataset}
\end{figure}

In our case, we assume that the location of each event instance is given by a pair of geographic coordinates (e.g., longitude and latitude). Distances between such coordinates can be converted to metric units using the Haversine formula, which allows $R$ to be expressed in meters or kilometers. Given the dataset of event instances $\mathbf{D}$, we obtain the following relationship between any pair of instances $i_1, i_2 \in \mathbf{D}$. In Fig.~\ref{fig:Dataset}, we showed an example dataset $\mathbf{D}$ with denoted following relationships between instances.  

\begin{definition}
Given a dataset of event instances $\mathbf{D}$, the corresponding set of event types $\mathbf{F}$, and the computed spatio-temporal following relationship $\prec$ on $\mathbf{D}$, we define a directed acyclic graph (DAG)
\[
G=(V,E),
\]
where nodes correspond to event instances and edges represent the spatio-temporal following relationships. Formally,
\[
V=\{\, i \mid i\in \mathbf{D}\,\},
\]
and for any two instances $i_1,i_2\in V$ we include a directed edge
\[
(i_1,i_2)\in E \quad \Longleftrightarrow \quad i_1 \prec i_2 .
\]
Each node $i\in V$ is labeled with its event type $i.F\in \mathbf{F}$.
\label{def:DAG}
\end{definition}

In Fig.~\ref{fig:Dataset}, we presented four DAGs $G_1, G_2, G_3, G_4$ of the event instances dataset $\mathbf{D}$ presented in this figure. 

For a given set of computed DAGs, we define an embedded sub-DAG as follows.

\begin{definition}
Given a DAG $G=(V,E)$, we define an \emph{embedded sub-DAG} as a graph $G'=(V',E')$ such that $V' \subseteq V$ and $E' \subseteq V' \times V'$. Moreover, for every edge $(u,v)\in E'$ there exists a directed path from $u$ to $v$ in $G$, i.e.,
\[
(u,v)\in E' \ \Longrightarrow\ u \rightarrow v \text{ in } G,
\]
where $u \rightarrow v$ denotes reachability in $G$ (not necessarily via a single edge).
\end{definition}

\begin{definition}
A frequent embedded sub-DAG is an embedded sub-DAG whose \emph{support} is at least $\mathit{minSup}$, where the support of a sub-DAG is defined as the fraction of DAGs in which it is embedded:
\[
\mathrm{sup}(G')=\frac{\left|\left\{\, G \in DAGs \;\middle|\; G' \text{ is embedded in } G \,\right\}\right|}{|DAGs|}.
\]
Here, $|DAGs|$ denotes the total number of DAGs in the collection, and $\mathit{minSup}\in(0,1]$ is a user-defined minimum support threshold.
\label{def:supp}
\end{definition}

Subsequently, we introduce the notion of a frequent closed embedded sub-DAG:
\begin{definition}
    A frequent embedded sub-DAG $G$ is said to be closed if there is no other frequent embedded sub-DAG $G'$ such that $G$ is embedded in $G'$ and support of both of them is the same. 
\end{definition}

By providing the user with only frequent closed embedded sub-DAGs, we can reduce the overall number of patterns presented to the user: from the set of all frequent closed embedded sub-DAGs a user can deduce the set of all frequent embedded sub-DAGs.

\begin{figure}
    \centering
    \includegraphics[width=0.75\textwidth]{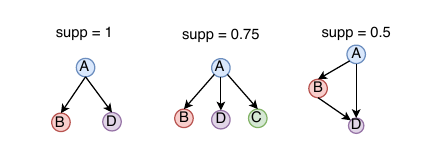}
    \caption{The set of all frequent closed embedded sub-DAGs for the dataset in Fig.~\ref{fig:Dataset}.} 
\label{fig:frequent}
\end{figure}

For four DAGs shown in Fig.~\ref{fig:Dataset} and for $minSup = 0.3$, the set of all frequent closed embedded sub-DAGs is presented in Fig.~\ref{fig:frequent}.

Thus, the task addressed in this work is to discover all frequent closed (embedded) sub-DAGs for a given input dataset $\mathbf{D}$ and parameters $R$, $T$, and $minSup$. For the data mining process, we employ the DigDag algorithm \cite{Termier2007-digdag}. We implemented this algorithm in Python using the \textit{networkx} library to represent DAGs.

Specifically, the procedure consists of the following steps. First, (i) we transform the dataset $\mathbf{D}$ of event instances into a set of DAGs. Subsequently, (ii) the DigDag algorithm is applied, which decomposes each DAG into a set of tiles and then performs closed frequent itemset mining to discover frequent patterns over the generated tiles. For each DAG obtained in step (i), a set of tiles is constructed by considering each non-leaf node as a root of a tile; each such node generates exactly one tile consisting of the node itself and the set of all its descendants. Overall, in our approach, we utilize the DigDag algorithm as an internal component, without introducing any modifications to its original design or operation.

\section{Experiments}

We conduct both quantitative and qualitative analysis of frequent closed embedded sub-DAGs mining. To conduct experiments, we used the \textit{Boston Crime Incidents Reports Dataset} provided by the Boston Police Department \cite{ref1284:BostonCrimeIncidents}. The dataset was collected between 08.07.2012 and 10.08.2015; however, only incidents from 01.01.2014 to 31.12.2014 were used in the experiments. For the experiments we selected attributes: \textit{incident location} (given by longitute and latitude), \textit{incident time} and \textit{incident type}. The dataset contains 26 crime event types, such as \textit{auto theft}, \textit{simple assault} or \textit{vandalism}. To calculate the spatial distance between event instances we used the Haversine formula.

\subsection{Quantitative Analysis}

\subsubsection{Comparison with Propagation Patterns (the SLEUTH algorithm)}

First, we compare the mining times of our approach with those of propagation patterns introduced in~\cite{Moosavi2019-ShortLongSTPatterns}. As discussed in Section~\ref{sec:RelatedWork}, propagation patterns are defined as frequent embedded subtrees and rely on a support definition analogous to Definition~\ref{def:supp}. Moreover, their discovery also requires the computation of the spatio-temporal following relationship between event instances.

In this set of experiments, we aim to evaluate the efficiency of both approaches under varying values of the $R$ and $T$ parameters, as well as for different $minSup$ thresholds. In general, increasing $R$ and $T$ reduces the number of constructed DAGs, while simultaneously increasing the number of nodes and edges within each DAG \footnote{For sufficiently small $R$, $T$, for each event instance, there is created a singleton DAG.}.

To discover propagation patterns, we employ the SLEUTH algorithm~\cite{Zaki2005-FrequentUnorderedTrees,Moosavi2019-ShortLongSTPatterns}, which requires the input DAGs to be directed trees (this limitation is discussed in Section~\ref{sec:RelatedWork}). Therefore, in our framework, each DAG constructed according to Definition~\ref{def:DAG} is transformed into an arborescence (i.e., a directed tree), where the root corresponds to the node representing the earliest event instance, in the following way.

For each node that has more than one parent in the original DAG, only a single parent is retained in the resulting arborescence. The parent is selected based on the earliest occurrence time. Consequently, SLEUTH operates on the same number of input graphs (trees) as DigDag; however, these graphs contain fewer edges due to the imposed tree structure \footnote{Please note that it is also possible to employ the DigDag mining to arborescences. However, in our approach we wanted to test the SLEUTH algorithm indicated by \cite{moosavi2019countrywide}.}.

\begin{figure}
    \centering
    \includegraphics[width=1\textwidth]{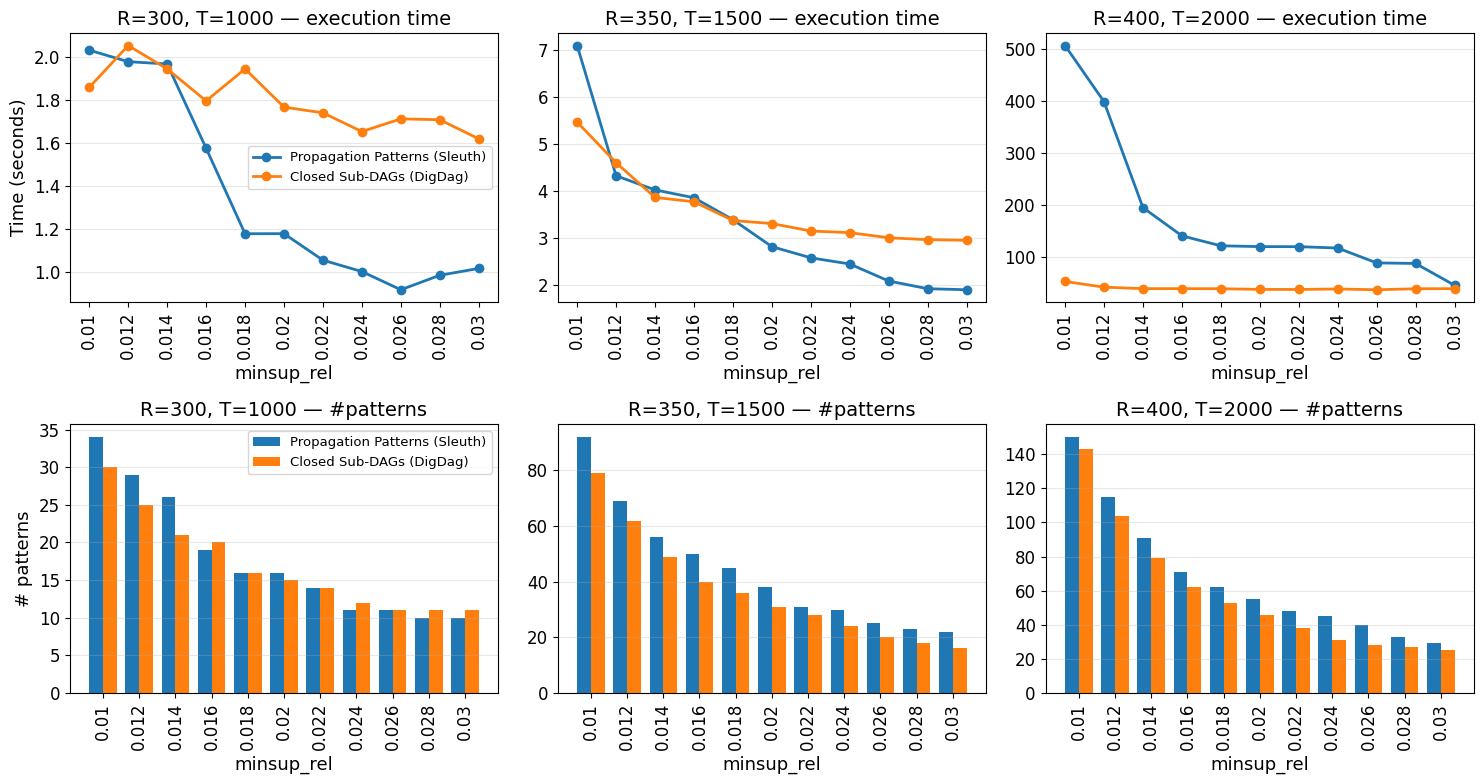}
    \caption{Execution times and numbers of discovered patterns for our approach and Propagation Patterns \cite{Moosavi2019-ShortLongSTPatterns} for different R, T and $minSup$ thresholds.} 
\label{fig:sleuth}
\end{figure}

The results of our experiments are presented in Fig.~\ref{fig:sleuth}. We conducted tests for three different configurations of the $R$ and $T$ parameters. It can be observed that for the first configuration ($R = 300$ meters, $T = 1000$ minutes), propagation patterns are discovered slightly faster than closed sub-DAGs. However, as $R$ and $T$ increase, the discovery of closed sub-DAGs using the DigDag algorithm becomes significantly more efficient. In particular, for $R = 400$ meters and $T = 2000$ minutes, DigDag executes considerably faster, and its execution times remain much more stable with respect to varying values of the $minSup$ parameter.

Since DigDag discovers only closed sub-DAGs, the number of returned patterns tends to be smaller than the number of Propagation Patterns, due to the elimination of redundant (non-closed) patterns. 

The difference in pattern discovery efficiency stems from the size of the explored search space. DigDag restricts mining to closed sub-DAGs, reducing the number of candidate patterns, whereas SLEUTH enumerates all frequent embedded tree patterns, requiring repeated support computation for many redundant candidates.


\subsubsection{Comparison with CSTPs patterns (the CSTPM algorithm)}

As mentioned in Section~\ref{sec:RelatedWork}, Mohan et al.~\cite{ref1284:Mohan2012} proposed the discovery of CSTP patterns, which are mined from a set of DAGs constructed based on spatio-temporal following relationships between event instances. However, they adopted a different algorithmic strategy for uncovering CSTPs. In each iteration, their approach generates new candidate patterns by extending existing CSTPs with additional nodes (corresponding to event types) or edges, and subsequently evaluates the significance of these candidates using the Cascading Participation Index (CPI) and Cascading Participation Ratios (CPRs).

Similarly to the $support$ measure used in our approach, which takes values in the interval $[0,1]$, the CPI measure is also defined within the same range, although the two measures are not strictly equivalent. Our implementation of the CSTPM algorithm was prepared according to~\cite{ref1284:Mohan2012}, and we implemented the two pruning strategies described therein the Upper Bounding (UB) filter and the Multiresolution Spatio-Temporal (MST) filter.
\begin{itemize}
    \item \textbf{UB filter:} Calculaties the ratio of the minimum to maximum CPR within a CSTP, giving an upper bound on $CPI(CSTP)$. Candidates below the threshold are pruned.

    \item \textbf{MST filter:} Computes CPI on a coarse dataset (spatial cells of size $d$, temporal intervals $t$). As it overestimates true CPI, candidates below the threshold are safely pruned early. In our experiments, we set the spatial cells size to $d = R \cdot 200$ and temporal intervals to $t = T \cdot 100$). 
\end{itemize}

\begin{figure}
    \centering
    \includegraphics[width=1\textwidth]{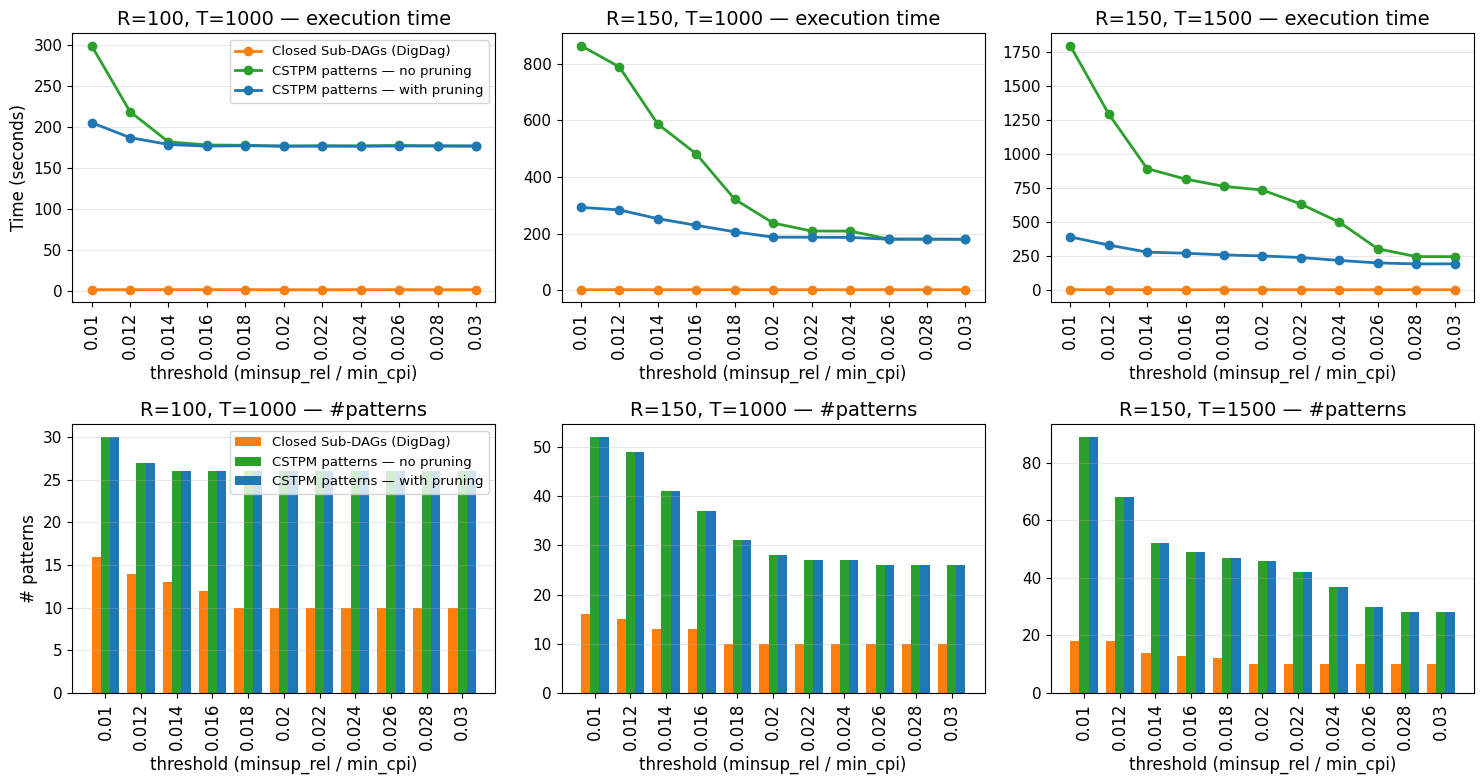}
    \caption{Execution times and numbers of discovered patterns for our approach and CSTPs \cite{ref1284:Mohan2012} for different R, T and $minSup$ and $minCPI$ thresholds.} 
\label{fig:cstpm}
\end{figure}

In Fig.~\ref{fig:cstpm}, we present a comparison of mining times and the numbers of discovered patterns for CSTPM (with and without pruning) and closed sub-DAGs (DigDag). In terms of execution times, DigDag is significantly faster. Similarly to the SLEUTH algorithm, CSTPM performs computationally expensive candidate pattern enumeration and CPI calculations. For lower CPI thresholds, this process becomes particularly costly due to the large number of generated candidates. The number of patterns discovered by CSTPM is slightly higher: this is resultant for a different definition of a significant CSTP pattern. Specifically, CSTPM classify all singleton patterns, each consisting of a single event type, as significant. 

\subsubsection{Discussion of Quantitative Analysis}

There are two important conclusions based on conducted quantitative analysis of the above results:

\begin{itemize}
    \item Discovering frequent closed embedded sub-DAGs using the DigDag algorithms is much faster than mining propagation patterns using the SLEUTH algorithm as proposed by Moosavi et al. \cite{Moosavi2019-ShortLongSTPatterns}. The execution times of DigDag for given support thresholds are also more stable that the execution times of SLEUTH. This is especially evident for more complex DAGs, for which execution times of DigDag significantly outperform those of SLEUTH.
    
    Furthermore, as we argued in Section~\ref{sec:RelatedWork}, due to the nature of spatio-temporal following relationships between event instances, discovering frequent sub-DAGs instead of frequent sub-Trees is better theoretically grounded. 

    \item Mining times with DigDag also significantly outperform mining times of the CSTPM algorithm for similar sets of input parameters used in both algorithms. The notion of CSTP patterns is closely related to the frequent sub-DAGs. Thus, in our opinion, a user should prefer discovering frequent sub-DAGs instead of CSTPs when mining spatio-temporal event data.
\end{itemize}

\subsection{Qualitative Analysis}

\begin{figure}
    \centering
    \includegraphics[width=1\textwidth]{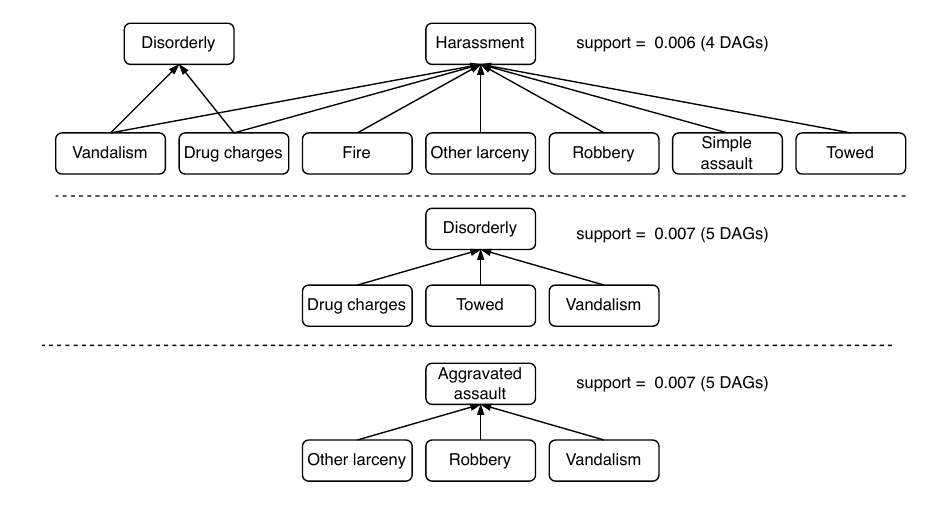}
    \caption{The examples of frequent closed sub-DAGs discovered from the dataset.} 
\label{fig:exp-patter}
\end{figure}

In this subsection, we closely look at the examples of some closed frequent sub-DAGs discovered in our approach. We first calculated spatio-temporal following relationships between event instances for parameters $R = 400$ meters and $T = 14400$ minutes (10 days). This resulted in 626 total input DAGs provided to the DigDag algorithm. We mined patterns with minimum support threshold $minSup = 0.005$ obtaining 1321 frequent patterns. After analysing them one-by-one, below we visualize some of them in Fig.~\ref{fig:exp-patter}. 

Based on the patterns presented in Fig.~\ref{fig:exp-patter}, we may observe, for example, that occurrences of events of type \textit{Disorderly} are in some cases preceded by simultaneous occurrences of events such as \textit{Drug Charges}, \textit{Towed}, and \textit{Vandalism}. Such a hypothesis, however, requires further investigation, particularly given the relatively large temporal window $T$ (10 days) used to determine the spatio-temporal following relationship.

In our opinion, patterns of the type illustrated in Fig.~\ref{fig:exp-patter} may play a \textit{supporting} role in the analysis of criminal activity in municipal areas. A competent authority should conduct further studies to assess the practical value of the discovered patterns, potentially combining our approach with other data mining techniques, such as clustering or classification.

\section{Conclusions}

In this paper, we proposed a novel approach for discovering frequent closed (embedded) sub-DAGs in spatio-temporal event data and introduced the necessary formal foundations for this class of patterns. We implemented the DigDag algorithm and experimentally demonstrated that our method is significantly more efficient than existing approaches for mining propagation patterns and CSTPs, namely SLEUTH and CSTPM. Additionally, we provided a qualitative analysis of selected discovered patterns, illustrating the practical usefulness and interpretability of the proposed framework.

\begin{credits}
\subsubsection{\ackname} The author was supported by the Warsaw University of Technology Research University - Excellence Initiative program [Grant number CPR-IDUB/ 288/Z01/POB3/2024].

\subsubsection{\discintname}
The author has no competing interests to declare that are
relevant to the content of this article.

\subsubsection{The use of AI tools}
The LLM model (ChatGPT-5) was used to polish writing of the manuscript, mainly including correction of grammar mistakes, and for supporting the preparation of the algorithms implementations used in the experiments.

\end{credits}

\bibliographystyle{splncs04}
\bibliography{my-references}

\end{document}